\documentclass[acmsmall]{acmart}

\usepackage[utf8]{inputenc}
\usepackage{algorithmic}
\usepackage{graphicx}
\usepackage{textcomp}
\usepackage{xcolor}
\usepackage[frozencache,cachedir=_minted-main]{minted}
\usepackage{subcaption}
\usepackage{hyperref}
\usepackage[capitalise]{cleveref}
\usepackage{booktabs}
\usepackage{csquotes}

\definecolor{Set1Red}{HTML}{e41a1c}
\definecolor{Set1Blue}{HTML}{377eb8}
\definecolor{Set1Green}{HTML}{4daf4a}
\definecolor{Set1Gray}{HTML}{beaed4}

\setcopyright{cc}
\setcctype{by}
\acmDOI{10.1145/3808178}
\acmYear{2026}
\acmJournal{PACMSE}
\acmVolume{3}
\acmNumber{FSE}
\acmArticle{FSE171}
\acmMonth{7}
\acmSubmissionID{fse26mainb-p1422-p}
\received{2025-09-12}
\received[accepted]{2026-03-24}

\begin{document}

\newcommand{\todo}[1]{\textcolor{red}{TODO: #1}}
\newcommand{\name}[1]{\textsc{PoCGen}}
\newcommand{\dsname}[1]{CWEBench.js}
\newcommand{\numSecbenchTotalUnfiltered}{600}
\newcommand{\numCWEBenchPastKnowledgeCutoff}{171}
\newcommand{\numSecBenchNoRefinements}{155}
\newcommand{\numSecBenchRefinements}{277}
\newcommand{\numSecBenchRefinementsLessThanTen}{221}
\newcommand{\numSecBenchTotal}{560}
\newcommand{\numSecBenchGHSA}{7}
\newcommand{\numSecBenchSnyk}{553}
\newcommand{\numSecBenchWorking}{432}
\newcommand{\numSecBenchFailures}{105}
\newcommand{\successRateSecBench}{71\%}
\newcommand{\timePerExploitSecBench}{11 minutes}
\newcommand{\timePerExploitSuccessSecBench}{7 minutes}
\newcommand{\timePerExploitCodeQLSecBench}{2 minutes}
\newcommand{\timePerExploitLLMSecBench}{5 minutes}
\newcommand{\completionTokensAverageSecBench}{17750}
\newcommand{\promptTokensAverageSecBench}{61234}
\newcommand{\completionTokensAverageSuccessSecBench}{8022}
\newcommand{\promptTokensAverageSuccessSecBench}{20701}
\newcommand{\completionTokensAverageFailureSecBench}{57111}
\newcommand{\promptTokensAverageFailureSecBench}{222675}
\newcommand{\costPerExploitSecBench}{0.02}
\newcommand{\costPerExploitSuccessSecBench}{0.008}
\newcommand{\costPerExploitFailureSecBench}{0.07}
\newcommand{\successRateSecBenchredos}{41\%}
\newcommand{\successRateSecBenchpathTraversal}{89\%}
\newcommand{\successRateSecBenchprototypePollution}{84\%}
\newcommand{\successRateSecBenchcodeInjection}{34\%}
\newcommand{\successRateSecBenchcommandInjection}{92\%}
\newcommand{\numSecBenchTaintPathSameFile}{404}
\newcommand{\numSecBenchTaintPathSameFunction}{277}
\newcommand{\numCWEBenchTotal}{794}
\newcommand{\numCWEBenchGHSA}{755}
\newcommand{\numCWEBenchSnyk}{39}
\newcommand{\numCWEBenchWorking}{312}
\newcommand{\numCWEBenchFailures}{423}
\newcommand{\successRateCWEBench}{39\%}
\newcommand{\timePerExploitCWEBench}{27 minutes}
\newcommand{\timePerExploitSuccessCWEBench}{24 minutes}
\newcommand{\timePerExploitCodeQLCWEBench}{3 minutes}
\newcommand{\timePerExploitLLMCWEBench}{6 minutes}
\newcommand{\completionTokensAverageCWEBench}{25900}
\newcommand{\promptTokensAverageCWEBench}{116067}
\newcommand{\completionTokensAverageSuccessCWEBench}{11274}
\newcommand{\promptTokensAverageSuccessCWEBench}{37426}
\newcommand{\completionTokensAverageFailureCWEBench}{37124}
\newcommand{\promptTokensAverageFailureCWEBench}{178581}
\newcommand{\costPerExploitCWEBench}{0.03}
\newcommand{\costPerExploitSuccessCWEBench}{0.012}
\newcommand{\costPerExploitFailureCWEBench}{0.05}
\newcommand{\successRateCWEBenchredos}{28\%}
\newcommand{\successRateCWEBenchpathTraversal}{32\%}
\newcommand{\successRateCWEBenchprototypePollution}{46\%}
\newcommand{\successRateCWEBenchcodeInjection}{18\%}
\newcommand{\successRateCWEBenchcommandInjection}{55\%}
\newcommand{\numCWEBenchTaintPathSameFile}{268}
\newcommand{\numCWEBenchTaintPathSameFunction}{153}
\newcommand{\numNpmInstallErrors}{39}
\newcommand{\code}[1]{\small\texttt{#1}\normalsize}
\newcommand\michael[1]{\textcolor{blue} {\footnotesize #1 -- {\em M}}}
\definecolor{darkgreen}{rgb}{0.1,0.6,0.0}
\newcommand{\aryaz}[1]{\textcolor{darkgreen} {\footnotesize #1 -- {\em A}}}

\setminted{
  fontsize=\footnotesize,
  linenos,
  numbersep=5pt,
  xleftmargin=10pt,
  frame=lines,
  breaklines=true,
  escapeinside=@@
}

\title{PoCGen: Generating Proof-of-Concept Exploits for Vulnerabilities in Npm Packages}

\author{Deniz Simsek}
\orcid{0009-0007-0799-2830}
\affiliation{%
  \institution{University of Stuttgart}
  \city{Stuttgart}
  \country{Germany}
}
\email{dsimk@pm.me}

\author{Aryaz Eghbali}
\orcid{0000-0001-9763-8147}
\affiliation{%
  \institution{CISPA Helmholtz Center for Information Security}
  \city{Stuttgart}
  \country{Germany}
}
\email{aryaz.egh@gmail.com}

\author{Michael Pradel}
\orcid{0000-0003-1623-498X}
\affiliation{%
  \institution{CISPA Helmholtz Center for Information Security}
  \city{Stuttgart}
  \country{Germany}
}
\email{michael@binaervarianz.de}

\setminted{fontsize=\footnotesize}

\begin{abstract}
  Security vulnerabilities in software packages are a significant concern for developers and users alike.
  Patching these vulnerabilities in a timely manner is crucial to restoring the integrity and security of software systems.
  However, previous work has shown that vulnerability reports often lack proof-of-concept (PoC) exploits, which are essential for fixing the vulnerability, testing patches, and avoiding regressions.
  Creating a PoC exploit is challenging because vulnerability reports are informal and often incomplete, and because it requires a detailed understanding of how inputs passed to potentially vulnerable APIs may reach security-relevant sinks.
  In this paper, we present \name{}, a novel approach to autonomously generate and validate PoC exploits for vulnerabilities in npm packages.
  The approach combines the complementary strengths of LLMs (e.g., understanding informal vulnerability reports), static analysis (e.g., identifying taint paths), and dynamic analysis (e.g., validating generated exploits).
  \name{} successfully generates exploits for 71\% of the vulnerabilities in the SecBench.js dataset.
  This success rate significantly outperforms a recent baseline (by 38 absolute percentage points), while imposing an average cost of only \$0.02 per generated exploit.
  Moreover, \name{} generates successful exploits for 60\% of 126 recent real-world vulnerabilities, which helped augment five recent vulnerability reports in the GitHub Security Advisories database with \name{}-generated PoC exploits.
\end{abstract}



\maketitle

\section{Introduction}

Security vulnerabilities pose a major threat to software and users alike, with the number of reported vulnerabilities increasing each year.
In 2024 alone, over 40,000 CVEs were disclosed, an increase of 38\% over the previous year~\cite{CVEs2024}.
As software ecosystems become more complex and interdependent, mitigating vulnerabilities becomes increasingly challenging.
This holds particularly for Node.js and its package manager, npm, which form the backbone of the JavaScript and TypeScript ecosystem.
With millions of packages and a dense network of dependencies, the npm ecosystem is susceptible to a wide range of security risks~\cite{Zimmermann2019}, including transitive vulnerabilities, where a single vulnerable package can propagate security risks across thousands of applications~\cite{Zahan2022weak}.

When a vulnerability is discovered, it is typically reported to the developers of the affected package, who are then expected to create a patch to fix the issue.
Once the vulnerable software is fixed, or some time has passed since the initial vulnerability report, the vulnerability report is published as a Common Vulnerabilities and Exposures (CVE) entry.
The process of fixing vulnerabilities is often facilitated by a proof-of-concept (PoC) exploit, which demonstrates how the vulnerability can be exploited in practice.
Moreover, PoC exploits are useful for testing the patch and preventing regressions in the future.
However, many vulnerability reports lack a PoC exploit~\cite{Householder2020}, and even many CVE reports do not have any PoC exploits.
For example, in the SecBench.js~\cite{Bhuiyan2023} dataset, only 179 out of \numSecBenchTotal{} CVEs contain a PoC exploit in the report.
Furthermore, the publicly available exploits are not reliable, and in some cases malicious themselves~\cite{Yadmani2023}.

\begin{figure}
  \begin{minted}[breaksymbol={}, breaklines=true]{markdown}
An issue in parse-uri v1.0.9 allows attackers to cause a Regular expression Denial of Service (ReDoS) via a crafted URL.
  \end{minted}
  \caption{CVE-2024-36751 report with no PoC exploit in the report.}
  \label{fig:cve_no_poc_example}
\end{figure}

As a real-world example, consider the vulnerability CVE-2024-36751 in the \enquote{parse-uri} package shown in \cref{fig:cve_no_poc_example}.
The vulnerability report informally describes the problem, but it does not contain an executable code example to reproduce it, i.e., there is no PoC exploit.
The process of creating PoC exploits is often time-consuming and requires a deep understanding of the codebase, the vulnerability, and the underlying technology~\cite{Bhuiyan2023}.
Particularly in the case of \cref{fig:cve_no_poc_example}, the vulnerability description does not mention which function is vulnerable, and does not provide any information about the input that triggers the vulnerability.

One way to address the challenge of generating a PoC exploit that is missing in a given vulnerability report is to leverage the capabilities of large language models (LLMs).
LLMs have demonstrated their effectiveness in various software engineering tasks, including code completion~\cite{Chen2021}, test generation~\cite{Schafer2024,Lemieux2023}, program repair~\cite{Jiang2023,Jin2023,Xia2024a,Zhang2024a,icse2025-RepairAgent}, and repository setup~\cite{issta2025_ExecutionAgent}.
They are especially good candidates because of their ability to understand natural language and other informal, imprecise, and vague information provided in vulnerability reports.
Moreover, with their understanding of different vulnerability types, LLMs can generate payloads for exploits in a targeted manner.
However, LLMs alone may not be sufficient to generate successful PoC exploits, as they lack the necessary context about the codebase, such as implementation details of the vulnerable code, and have only a limited ability to reason about the behavior of the code.

Another way to generate PoC exploits is to deploy traditional testing and analysis techniques, such as fuzzing~\cite{Cassel2025} and symbolic execution~\cite{Marques2025}.
The currently most effective such technique is a recent approach called \emph{Explode.js}~\cite{Marques2025}, which uses taint analysis, a set of exploit templates, and symbolic execution to produce exploits that trigger vulnerabilities in npm packages.
Explode.js has been shown to generate PoC exploits for 182 out of \numSecBenchTotal{} vulnerabilities (32\%) in the SecBench.js dataset, which is helpful, but still leaves many vulnerabilities without a PoC exploit.
We attribute this gap to the fact that Explode.js uses only traditional program testing and analysis techniques, leveraging neither the information provided in vulnerability reports nor the capabilities of LLMs.

This paper addresses these limitations by presenting \name{}, a novel approach for generating PoC exploits from informal vulnerability reports.

\name{} takes as input an informal description of a vulnerability, as found in CVE reports, as well as the vulnerable code base, and automatically generates an executable JavaScript file that demonstrates how to exploit the vulnerability.
The approach consists of four iteratively executed components: (i) understanding the vulnerability and extracting source-level information, (ii) generating a candidate exploit, (iii) validating the exploit, and if necessary, (iv) refining the prompt to obtain an improved exploit.
These components use a combination of LLM prompting and static and dynamic analysis techniques:
Component (i) uses dynamic analysis to explore the package's exported functions, prompts the LLM with the given vulnerability report to identify potentially vulnerable functions, and queries a static taint analysis to extract taint paths.
Component (ii) generates a candidate exploit using an LLM.
Component (iii) executes and validates the candidate exploit against a dynamic analysis-based test oracle.
If the candidate exploit is not valid, \name{} uses component (iv) and refines the prompt using a set of refiners that provide static or dynamic information to component (ii) where the LLM attempts again to generate a valid exploit.
This process continues until either a valid exploit is generated or until exceeding the computational budget.

To generate a PoC exploit for the ReDoS vulnerability in \cref{fig:cve_no_poc_example}, the main challenge is to construct a payload that, when passed to the vulnerable function, triggers a security-relevant action.
The goal of ReDoS is to exploit the regular expression engine's backtracking behavior, leading to excessive resource consumption.
After analyzing the codebase for potential vulnerable functions, \name{} identifies the package's default exported function as the likely vulnerable entry point, and extracts usage examples from the codebase.
In its first attempt, \name{} generates an exploit that passes a crafted input string designed to trigger the ReDoS vulnerability.
However, this payload is invalid as it does not cause the expected backtracking behavior.
Once \name{} executes the initial candidate exploit, it determines that the exploit fails and, reasoning about the vulnerability, the code, and the runtime information, generates a new payload.
After multiple refinements, \name{} generates a new exploit that triggers the ReDoS vulnerability, as shown in \cref{fig:poc_exploit_ex1}.

\begin{figure}
  \begin{minted}{javascript}
async function exploit() {
    const parseuri = require("parse-uri");
    // This input is designed to cause excessive backtracking in the regex
    const craftedInput = 'http://example.com/' + 'a'.repeat(30000) + '?key=value';
    const result = await parseuri(craftedInput);
}
await exploit();
\end{minted}
\caption{\name{}-generated PoC exploit for CVE-2024-36751.}
\label{fig:poc_exploit_ex1}
\end{figure}

We evaluate \name{} on SecBench.js, a benchmark of vulnerable npm packages with path traversal, prototype pollution, command injection, code injection, and ReDoS vulnerabilities.
Our results show that \name{} successfully generates exploits for 395 out of \numSecBenchTotal{} vulnerabilities (71\%), outperforming the previous state of the art~\cite{Marques2025} by 38 absolute percentage points.
When measuring the costs due to LLM usage, we find that \name{} incurs an average cost of only \$\costPerExploitSecBench{} per vulnerability, which is reasonably low for adoption in practice.
To further validate the practical usefulness of \name{}, we use it to generate PoC exploits for 126 recent real-world vulnerabilities, where \name{} successfully generates PoC exploits for 76 of them (60.3\%).
Five of the generated exploits have been added to the vulnerability reports on the GitHub Security Advisories database.

We envision the approach to be useful for developers and security researchers.
By using \name{}, developers of npm packages can generate PoC exploits for vulnerability reports they receive to help them understand the vulnerability and how to address it.
They can also use the PoC exploits to test their patches and even add them to their test suites as regression tests.
Moreover, security researchers can automate reporting vulnerabilities to downstream packages, by automating the PoC exploit generation process.

In summary, this paper makes the following contributions:
\begin{itemize}
  \item A novel approach to autonomously generate and validate PoC exploits for vulnerabilities in npm packages by combining LLMs with static and dynamic analysis.
  \item Empirical evidence of the effectiveness of \name{} on \numSecBenchTotal{} real-world vulnerabilities, covering five common vulnerability types, as well as its utility for augmenting vulnerability reports that were previously missing PoC exploits.
  \item We share our code and data to foster future research (Section~\ref{sec:data}).
\end{itemize}

\section{Approach}

\subsection{Overview}

\begin{figure*}
  \centering
  \includegraphics[width=\linewidth]{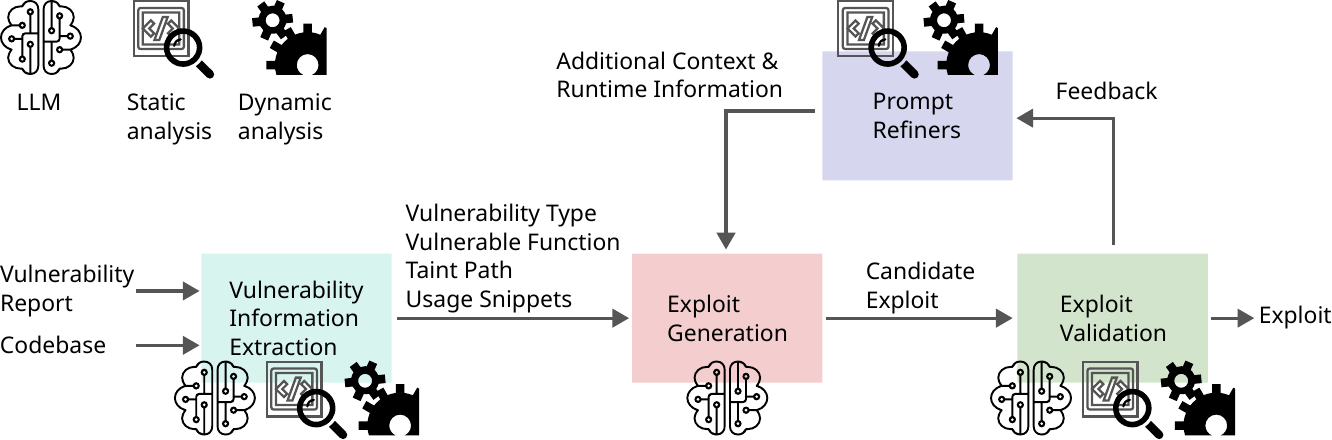}
  \caption{Overview of \name{}.}
  \label{fig:overview}
\end{figure*}
\Cref{fig:overview} provides an overview of \name{}.
The approach consists of four main components, which combine LLM prompting, static analysis, and dynamic analysis.
\name{} takes as input a vulnerability report, which is an informal description of the vulnerability, and the codebase of the vulnerable package. 
Vulnerability reports originate from several sources, such as vulnerability databases (e.g., the CVE database, GitHub Security Advisories, and Snyk), bug and issue trackers, or security mailing lists.

The first step of \name{} (Section~\ref{sec:vuln_info_extraction}) is to extract information about the vulnerability and the codebase.
This step is necessary because the natural language description in a vulnerability report is often vague and does not provide enough information to directly generate a PoC exploit.
The approach extracts the vulnerability type, the likely vulnerable function, taint paths to vulnerable sinks, and usage examples.

In the second step (Section~\ref{sec:exploit_generation}), \name{} uses the extracted information to generate a candidate exploit.
To this end, the approach compiles all the available information into a prompt that asks the LLM to generate a PoC exploit.

Once the LLM generates a candidate exploit, the third step of \name{} (Section~\ref{sec:exploit_validation}) executes and validates the candidate exploit.
The approach uses multiple validation mechanisms, including a set of runtime checkers that are specific to each vulnerability type.
For example, for command injection vulnerabilities, the validation checks whether a specific command is executed after running the candidate exploit, while for prototype pollution vulnerabilities, it checks whether a specific property is added to the global object.

If the candidate exploit is successfully validated, \name{} returns it to the user and the approach terminates.
Otherwise, the fourth step of the approach (Section~\ref{sec:prompt_refinement}) refines the exploit generation prompt by adding additional context based on static analysis and the runtime information obtained during exploit generation.
This process continues until either finding a valid exploit or exhausting the computational budget.

\subsection{Vulnerability Information Extraction}
\label{sec:vuln_info_extraction}

In this component, \name{} extracts four pieces of information from the vulnerability report and the codebase to provide as context to the exploit generation component.

\subsubsection{Vulnerability Type}
First, \name{} identifies the type of vulnerability.
This is crucial to our approach as the type of vulnerability determines the goal of the exploit and how it should be validated.
To do this, we prompt the LLM with the vulnerability report and ask it to identify the type of vulnerability.
As the vulnerability report is written in natural language, and the description is informal, an LLM is a suitable tool to extract information from it.
We provide the LLM with the five vulnerability types supported by our approach: path traversal, prototype pollution, command injection, code injection, and Regular Expression Denial of Service (ReDoS).
\name{} prompts the LLM to select the most relevant vulnerability type from the list.

In addition to the vulnerability type, the LLM also determines if the vulnerability is exploitable remotely, such as a vulnerability in an http server that requires triggering the vulnerability via a request, or locally, where the vulnerability is triggered by importing the package and calling a function.

\subsubsection{Vulnerable Function}
In our approach and throughout this paper, we refer to the function that is accessible to an attacker and is the entry point for the exploit as the \emph{vulnerable function}.
Finding the vulnerable function is helpful for generating the PoC exploit, as it provides information about the input types and possible values, and potentially the vulnerable execution path of the function.
To identify the vulnerable function, we first load the package and dynamically extract all the functions that it exports.
We then prompt the LLM with the vulnerability report and ask it to identify the vulnerable function from the extracted functions.
Since the vulnerability report can be vague, and multiple functions can be candidates for the vulnerable function, we prompt the LLM to rank the functions based on their likelihood of being the vulnerable function.
This ranked list of candidate functions is the input to the next step which is to extract taint paths.

\subsubsection{Taint Path Extraction}
\label{sec:taint_path}

To generate a successful PoC exploit, it is crucial to understand whether and how the input to the vulnerable function flows through the code and reaches the sensitive operations, commonly referred to as vulnerable sinks.
Taint analysis is a technique that tracks the flow of data through the code by marking a certain input as tainted and propagating this taint through the code.
Moreover, to narrow down the search space for the vulnerable function, \name{} attempts to extract a taint path from each of the candidate functions identified in the previous step to a vulnerable sink for locally exploitable vulnerabilities.
If no taint path is found for a candidate function, that function is not considered for exploit generation.
For remotely exploitable vulnerabilities, we also run the taint analysis starting at remote access endpoints, such as http request handlers.

In our approach we define a taint path as a sequence of source code lines starting with the signature line in the definition of the vulnerable function, and ending with a vulnerable sink.
Any lines of code that propagate the taint are also included in the taint path.
To provide more context to the LLM, for each line in the taint path, we include three lines before and after it as additional context.
If these context windows overlap, we merge them to avoid duplication.

We use the static taint analysis provided by CodeQL\footnote{\url{https://codeql.github.com/}}.
For each vulnerability type, we use the vulnerable sinks and taint propagation rules specified in the JavaScript security library of CodeQL\footnote{The module \texttt{semmle.javascript.security}}.

\name{} processes the ranked candidate functions in batches of 50, querying the taint analysis for each batch to extract at least one taint path; this also serves to pinpoint the vulnerable function.
The taint analysis of CodeQL is designed for industry-level security analysis, which requires high precision and few false positives.
This means that the taint analysis may not find all taint paths.

Hence, if the first taint tracking attempt does not find any taint paths, our approach retries the taint analysis with our own extended set of taint propagation rules and vulnerable sinks.
If the extended taint analysis is also unsuccessful, \name{} prompts the LLM to guess the vulnerable sinks, and if that also fails, it falls back to a combination of static and dynamic taint analysis.
Concretely (anticipating Section~\ref{sec:exploit_generation}), it first prompts the LLM to generate an initial exploit for the vulnerable function, then executes that exploit and runs the static taint analysis on the functions called during execution.
If any of the executed functions have a taint path to a vulnerable sink, the approach uses this taint path as the taint path for the vulnerable function.

If \name{} still does not find a taint path, it proceeds to the next batch of candidate functions from the ranking and repeats the process.
If the approach exhausts all candidate functions without finding any taint paths, it considers the exploit generation attempt as failed and does not move to further steps of the approach.

The output of this step is a sequence of code snippets interleaved with natural language explanations of how the code propagates the taint, shown in \cref{fig:taint_path}, with multiple sections if the taint path spans multiple files.
In each section, there is a header specifying the file, followed by the taint path and its additional context from that file.
The taint path lines are also marked by a comment at the end of the line.

The package \code{djv} in \cref{fig:taint_path}, which is a JSON schema validator, had a code injection vulnerability (CVE-2020-28464).
The vulnerable function \code{import} takes a JSON string as input, parses it, and then calls the \code{restore} function with each of the attributes of the parsed object.
In the restore function, a new function is dynamically created where the body of the function comes from the parsed JSON string.
As shown in \cref{fig:taint_path}, a malicious input (e.g., JSON string containing arbitrary JavaScript code) to the \code{import} function, flows through to the \code{restore} function, and is used as the body of a dynamically created function, which is a code injection vulnerability.

\begin{figure}
  \begin{minted}{markdown}
Vulnerable method `import` of class `Environment` located in djv/lib/djv.js:
```js
import(config) { // tainted: "config"
  const item=JSON.parse(config) // tainted: "config"
  let restoreData=item // tainted: "item"
  if (item.name && item.fn && item.schema) {
    restoreData={
      [item.name]: item,
    }
  }
  Object.keys(restoreData).forEach((key)=>{ // tainted: "restoreData"
    const {name, schema, fn: source}=restoreData[key] // tainted: "restoreData"
    const fn=restore(source, schema, this.options) // tainted: "source"
    this.resolved[name]={
      name,
      schema,
```
Call to `restore`:
```js
function restore(source, schema, {inner}={}) { // tainted: "source"
  const tpl=new Function("schema", source)(schema) // tainted: "source"
  if (!inner) {
```
  \end{minted}
  \caption{Example of a taint path extracted by \name{} from the package \code{djv} with code injection vulnerability (CVE-2020-28464).}
  \label{fig:taint_path}
\end{figure}

\subsubsection{Usage Snippets}
\label{sec:usage_snippets}
To help the LLM generate a valid exploit, \name{} extracts usage examples of the vulnerable function from the codebase.
These examples allow the LLM to understand the function signature, and any setup that is required to call the function.
We extract usage snippets both from the source code and from the documentation of the package.
The usage snippets from the source code are extracted from test files using static analysis, by finding all the call sites of the vulnerable function.
For the usage snippets from the documentation, we first extract all code pieces in the documentation, wrapped in triple backticks ("\verb|```|"), and then we prompt an LLM to determine if they are usage examples of the vulnerable function.
If they are, we also prompt the LLM to summarize them.

\begin{figure}
\begin{minted}{javascript}
  const path = require('doc-path');
  let document = {
      Make: 'Nissan',
      Features: [ { feature: 'A/C' } ]
  };
  console.log(path.setPath(document, 'Color.Interior', 'Tan'));
  /* {
          Make: 'Nissan',
          Features: [ { feature: 'A/C' } ]
          Color: { Interior: 'Tan' }
      } */
  console.log(path.setPath(document, 'StockNumber', '34567'));
  /* {
          Make: 'Nissan',
          Features: [ { feature: 'A/C' } ]
          Color: { Interior: 'Tan' },
          StockNumber: '34567'
      } */
  console.log(path.setPath(document, 'Features.cost', '$0 (Standard)'));
  /* {
          Make: 'Nissan',
          Features: [ { feature: 'A/C', cost: '$0 (Standard)' } ]
          Color: { Interior: 'Tan' },
          StockNumber: '34567'
      } */
\end{minted}
\caption{Example of extracted usage snippets for the vulnerable function \code{setPath} in the \code{doc-path} package.}
\label{fig:usage_snippets}
\end{figure}

\subsection{Exploit Generation}
\label{sec:exploit_generation}
\name{} assembles a prompt for the LLM to generate the PoC exploit.
\Cref{fig:first_prompt} shows the prompt template that we use for generating the exploit.
The prompt starts with naming the vulnerable function (as \code{vulnerableFunction}) and the vulnerability type (as \code{vulnerabilityType}), which are extracted in the previous step, followed by a description of the vulnerability (as \code{vulnerabilityDescription}), which is the text of the vulnerability report.
Then, we provide the example usages of the vulnerable function (as \code{usageSnippets}), which are extracted from the codebase in the previous step.

Next, the prompt describes what the generated exploit should look like by providing the skeleton of the exploit code (as \code{exploitSkeleton}), and also providing exploits of similar vulnerabilities (as \code{similarExploits}).
The skeleton of the exploit code is a fixed template that contains the definition of the exploit function, its call, and loading the vulnerable function from the vulnerable package.
To extract similar exploits, our approach uses BM25 to find the three most similar vulnerability descriptions from the SecBench.js dataset to the current vulnerability report, and provides their exploits as examples.

After providing the context, the prompt specifies the task to the LLM, which is to create an exploit that uses the vulnerability in the vulnerable function to trigger a security-relevant action (as \code{vulnerabilityTrigger}).
This action is specifically defined for each vulnerability type, to facilitate the validation of the generated exploit.
Finally, the prompt provides the taint path snippets (as \code{taintPathSnippets}), which are extracted in the previous step.

\begin{figure}
  \centering
  \begin{minted}[breaksymbol={}]{handlebars}
`{{vulnerableFunction}}` is vulnerable to {{vulnerabilityType}}.
## Vulnerability Description:
```
{{vulnerabilityDescription}}
```
{{> usageSnippets }}
{{> exploitSkeleton }}
## Task:
Create an exploit that uses the {{vulnerabilityType}} in `{{vulnerableFunction}}` to {{vulnerabilityTrigger}}.
Respond with the full exploit code and explain why it works.
If there is an exception thrown, do not try to handle it and pass it on.
Enclose the exploit code in backticks and define the exploit within a function named `exploit`.
{{> similarExploits }}
## Source code:
{{taintPathSnippets}}
  \end{minted}
  \caption{Prompt template for the first round of exploit generation.}
  \label{fig:first_prompt}
\end{figure}

\subsection{Exploit Validation}
\label{sec:exploit_validation}

For each vulnerability type, \name{} uses a specific validation mechanism to check whether the generated exploit is valid.
These mechanisms are inspired by the validation checks in SecBench.js~\cite{Bhuiyan2023}.
Each validator is designed to check for an event that is commonly considered as the effect of a successful exploit for the corresponding vulnerability type.
For example, for the path traversal vulnerability, if the exploit is able to access a flag file in one or more directories above the current directory (e.g., \code{../../flag.txt}), the exploit is considered valid.
Our prompts force the model to generate exploits that trigger these specific events, which makes the validation process more straightforward and reliable.
Therefore, passing the validator is a requirement for an exploit to be considered valid.
This means that, by definition, we do not have any false negatives, as any exploit that does not trigger the expected event is considered to be unable to exploit the vulnerability.
We explain the checks each validator performs for each vulnerability type in the following.

\subsubsection{Path Traversal}
The goal of path traversal exploits is to access a file in the root directory.
In the validators of SecBench.js this was implemented as being able to read a flag file in the root directory.
However, this would render exploits for vulnerabilities that perform other file operations invalid.
To mitigate this, we consider all file system operations during runtime, and check the normalized path of the accessed file.
If the normalized path is evaluated to \code{flag.txt} in a directory above the current directory, the exploit is considered valid.

\subsubsection{Prototype Pollution}
In the case of prototype pollution vulnerabilities, the goal of the exploit is to add a property named \code{exploited} to the global \code{Object.prototype} object.
However, the property might be set by a direct assignment in the exploit code rather than through the vulnerable function, which would not constitute a genuine exploit.
To filter out such cases, we check whether the exploit code contains a direct assignment of the form \code{*[\_\_proto\_\_|prototype].exploited = *}.
If such an assignment exists and the property is set, the exploit is considered invalid.

\subsubsection{Command Injection}
For command injection vulnerabilities, the goal of the exploit is to execute the command \code{/usr/bin/genpoc}.
This command writes a file to the file system, which is then checked by the validation checker.

\subsubsection{Code Injection}
For code injection vulnerabilities, the goal of the exploit is to call the built-in \code{process.seteuid} function, with parameter 42.
This indicates that the exploit was able to access the \code{process} module.
Access to this module implies the ability to execute arbitrary code in the context of the Node.js process.
To ensure that the exploit triggers the sink through the vulnerable function rather than calling it directly, we statically check whether \code{process.seteuid(42)} is called directly in the exploit code.
If there is such a direct call, we consider the exploit invalid, as it bypasses the vulnerable function.

\subsubsection{ReDoS}
For ReDoS vulnerabilities, the goal of the exploit is to cause a denial of service by taking a long time to execute.
We hook the string and regex functions in the V8 engine to measure the time these functions take.
If the execution time of a function exceeds 1,500 milliseconds, we consider the exploit as valid.

\subsubsection{Sanity Checks}
Passing the vulnerability-specific validation checks is a necessary condition for an exploit to be considered valid, but it is not sufficient as it is possible to achieve the desired outcome with other means.
For example, in case of ReDoS, the LLM might generate a new regular expression that takes a long time to execute, but is not related to the vulnerable function.
Therefore, in addition to the vulnerability-specific validation checks, \name{} searches the call stack when the vulnerability is triggered for a call to the vulnerable function.
If no such call is found, the exploit is considered invalid.

\medskip
As a last step in the validation process of any vulnerability type, we prompt the LLM to check whether the exploit actually triggers the vulnerability described in the report.
This is done to filter out any invalid exploits that passed the previous validation checks.

\subsection{Prompt Refinement}
\label{sec:prompt_refinement}

After the validation step, if the exploit is not valid, \name{} refines the prompt to generate a new candidate exploit.
\name{} uses a set of refiners that provide static or dynamic information to the LLM to help it generate a valid exploit.

\subsubsection{Context Refiners}
The first set of refiners are the \emph{context refiners}, which provide additional context to the LLM to help it generate a valid exploit.
Since the taint path only contains the taint propagation lines, checks on taint values are not included in the taint path.
Therefore, the LLM might lack information about input-validation guards in the code that could prevent a malicious input from reaching the sink.
To address this, we provide a \emph{body refiner}, which provides the full body of any function that has at least one line of code in the taint path.

However, there can be checks that happen via function calls that are not in the taint path.
To address this, we also provide a \emph{missing declaration refiner}, which provides the LLM with the ability to ask for definitions of variables and functions in the taint path, through the function calling format of OpenAI's API.
The LLM can output a list of identifiers for which it needs their definitions, and \name{} will provide the definitions of these identifiers in the prompt using the V8's API, which itself uses a combination of static and dynamic analyses.

\subsubsection{Runtime Refiners}
\label{sec:runtime_refiners}
The second set of refiners are \emph{runtime refiners}, which add information about the execution to the prompt.
The refiners in this category are the \emph{error refiner}, the \emph{coverage refiner}, the \emph{debugger refiner}, and the vulnerability-specific refiners.

Since the exploit generated by the LLM can have runtime errors, for example from a wrong API usage, the \emph{error refiner} provides the LLM with the error message that was thrown during the execution of the candidate exploit.

The \emph{coverage refiner} provides the LLM with the coverage information of the candidate exploit, as markings in comments at the end of each line in the taint path.
This information is useful for the LLM to understand which parts of the code were not executed.
If the vulnerable sink was not executed, the information provided by this refiner can help the LLM to generate a new exploit that reaches the vulnerable sink.

The \emph{debugger refiner} provides the LLM with a debugger-like tool.
The LLM can output a list of expressions, for which it needs the runtime values.
Using the Inspector API of Node.js\footnote{\url{https://nodejs.org/api/inspector.html}}, \name{} evaluates these expressions during the execution of the exploit.
These values are provided as comments in their respective lines of the taint path in the prompt.

There are cases where the LLM generates an exploit that reaches the vulnerable sink, but the exploit fails the validation checks due to a wrong input.
For path traversal, command injection, and code injection vulnerabilities, we provide specific refiners that hook into the vulnerable sinks and provide the runtime values passed to these functions.
This form of feedback allows the LLM to understand how the input it generated is transformed, which can help it generate a valid exploit in the next iteration.
For path traversal vulnerabilities, the refiner provides the values passed to the file system functions, like \code{fs.readFile} and \code{fs.open}.
For command injection vulnerabilities, it provides the values passed to the \code{spawn} function.
Finally, in case of code injection vulnerabilities, the refiner provides the values passed to the most common sink functions, like the \code{Function} constructor.

\subsubsection{Combining Refiners}

\begin{figure}
  \begin{minted}{text}
prompts @$\leftarrow$@ PriorityQueue{(0, getSeedPrompt())}
usedRefiners @$\leftarrow$@ {}
seenExploits @$\leftarrow$@ {}
while prompts @$\neq$@ {} and |usedRefiners| < 30 do
    prompt @$\leftarrow$@ top(prompts)
    prompts @$\leftarrow$@ prompts \ {prompt}
    prompts @$\leftarrow$@ prompts @$\cup$@ {prompts with more info}
    exploit @$\leftarrow$@ generateExploit(prompt)      // via LLM
    if exploit @$\notin$@ seenExploits then
        result @$\leftarrow$@ runAndValidate(exploit)
        if result.success then
            return exploit
        endif
        correctlyUsedInfo @$\leftarrow$@ getCorrectlyUsedInfo(prompt, result) // Information in the prompt that the LLM used correctly
        prompts @$\leftarrow$@ prompts @$\cup$@ {(
          result.taintSteps + result.newErrors, 
          prompt + result.errors - correctlyUsedInfo
        )}
        usedRefiners @$\leftarrow$@ usedRefiners @$\cup$@ {used refiners}
        seenExploits @$\leftarrow$@ seenExploits @$\cup$@ {exploit}
    endif
endwhile
return failure
  \end{minted}
  \caption{Algorithm for the refinement loop.}
  \label{fig:refinement_loop}
\end{figure}

Based on the different refiners described above, \name{} implements a refinement loop that iteratively improves prompts until a valid exploit is generated or the budget is exhausted.
\Cref{fig:refinement_loop} shows the algorithm for the refinement loop.
Initially, a seed prompt is placed in the priority queue.
In every refinement attempt, \name{} chooses one refiner from the front of this queue.
Each time the approach uses a refiner, it assigns a score based on the number of new errors the respective exploit causes, and the number of steps from the taint path it covers.
It then adds the refined prompt to the priority queue with the score.
Moreover, if a refinement generates a prompt that is already used, \name{} does not query the LLM again, and moves to the next refiner.
To keep the prompts concise, in each refinement, \name{} removes parts of the prompt that the LLM has correctly used in the previous attempts.
For example, if the exploit generated in the previous step uses the vulnerable function correctly, \name{} removes the usage snippets from the prompt.
The refinement process continues until either a valid exploit is generated, or the computational budget is exhausted.

\section{Evaluation}
We evaluate \name{} on \numSecBenchTotal{} vulnerabilities in npm packages to answer the following research questions:
\begin{itemize}
  \item RQ1: How effective is \name{} in generating PoC exploits?
  \item RQ2: How much does each component of \name{} contribute to the overall effectiveness?
  \item RQ3: How much does it cost to generate PoC exploits in terms of money and time?
  \item RQ4: How effective is \name{} in recent real-world vulnerability reports?
  \item RQ5: What are the characteristics of vulnerabilities that affect the success of \name{}?
\end{itemize}

\subsection{Experimental Setup}

\subsubsection{Dataset}
We use the SecBench.js dataset~\cite{Bhuiyan2023} to evaluate \name{}
The dataset contains \numSecbenchTotalUnfiltered{} vulnerable npm packages with code injection, command injection, prototype pollution, path traversal, and ReDoS vulnerabilities.
We exclude packages that have been removed from the npm registry.
This leaves us with a total of \numSecBenchTotal{} vulnerabilities to evaluate our approach.
The vulnerable packages in SecBench.js are packages with various sizes spanning from 1 line of code to 892k lines of code, with the largest package being “jointjs”.
The average package has 8.3k lines of code.
Moreover, the vulnerabilities in these packages are real-world vulnerabilities and the average package in this benchmark is depended on by 947.62 packages, meaning they are used in many other JavaScript projects.

\subsubsection{Implementation and Configuration}
We run all experiments on an Ubuntu 22.04 machine with Intel Xeon(R) Silver 4214 CPU, with 256 GB of RAM.
The experimental setup uses Node.js version 22.11.0, running on a modified V8 engine that throws an error if a configurable backtracking limit is exceeded.
For static taint analysis, we use CodeQL version 2.20.4.

As the LLM, we use OpenAI's \texttt{gpt-4o-mini-2024-07-18} and \texttt{gpt-5-mini-2025-08-07} models through the OpenAI API.
We use the \texttt{gpt-5-mini} model with ``minimal'' reasoning effort as the default model for all experiments, unless otherwise specified.
We use a system prompt that assigns the role of a security researcher specialized in creating exploits for the identified security class to the LLM.
This is done to reduce refusals to generate exploits by the LLM for safety reasons.
For each vulnerability, we allocate a time budget of one hour, a token budget of 300k input tokens, and 100k output tokens.
The maximum refinement budget is set to 30 iterations.

\subsubsection{Baselines}

We compare \name{} against two baselines: Explode.js~\cite{Marques2025} and an LLM-based agent using the Mini-SWE-Agent framework\footnote{\url{https://mini-swe-agent.com}}~\cite{Yang2024a}.
Explode.js is a state-of-the-art approach for generating PoC exploits.
It first uses a static dataflow analysis to detect which exported functions reach a vulnerable sink, which is then used to create an exploit template.
Then, using symbolic execution, it generates symbolic inputs to exploit the vulnerability.
Finally, it uses an SMT solver to generate concrete inputs that trigger the vulnerability.

Recently, LLM-based agents have shown great promise in software engineering tasks, such as resolving issues~\cite{Yang2024a}, repairing bugs~\cite{icse2025-RepairAgent}, and executing arbitrary projects~\cite{issta2025_ExecutionAgent}.
Therefore, we also implement an LLM-based agent using the Mini-SWE-Agent framework as a second baseline.
Our version of the Mini-SWE-Agent is configured to target the PoC exploit generation task, and is provided with the same information as \name{}, i.e., the vulnerability report without any PoC exploits.
We limit the agent to run for at most 30 steps, similar to our approach.
The Mini-SWE-Agent can use the command line (bash) to perform various tasks, such as navigating the file system, reading and writing files, and executing code.

\subsection{RQ1: Effectiveness}
\label{sec:effectiveness}
We evaluate the effectiveness of \name{} in generating PoC exploits for vulnerabilities in SecBench.js as done in previous work~\cite{Marques2025}.
We measure the success rate of our approach and compare it to the baselines.
We also report the number of failed attempts, and the number of false positives, which are exploits that pass the validator, but do not trigger the vulnerability through the vulnerable function.
For each PoC exploit that \name{} generates, one of the authors manually inspects it to determine whether it is a false positive or a successful exploit.

In our experiments, we run \name{} three times on all \numSecBenchTotal{} vulnerabilities in SecBench.js to mitigate the randomness of LLMs.
We use the reported results of Explode.js\footnote{\url{https://github.com/formalsec/explode-js/blob/71ec17fe90f29e236b01b8cad02685344f8aff10/bench/explode-vulcan-secbench-results.csv}} on the same set of vulnerabilities.
However, since agentic approaches require evaluating their trajectory to ensure correctness, which is expensive and time-consuming, we limit the number of vulnerabilities evaluated with our agentic baseline approach to 100.
We randomly sample 20 vulnerabilities from each of the five vulnerability types in SecBench.js, and run Mini-SWE-Agent on them.

\begin{figure*}
  \includegraphics[width=\textwidth]{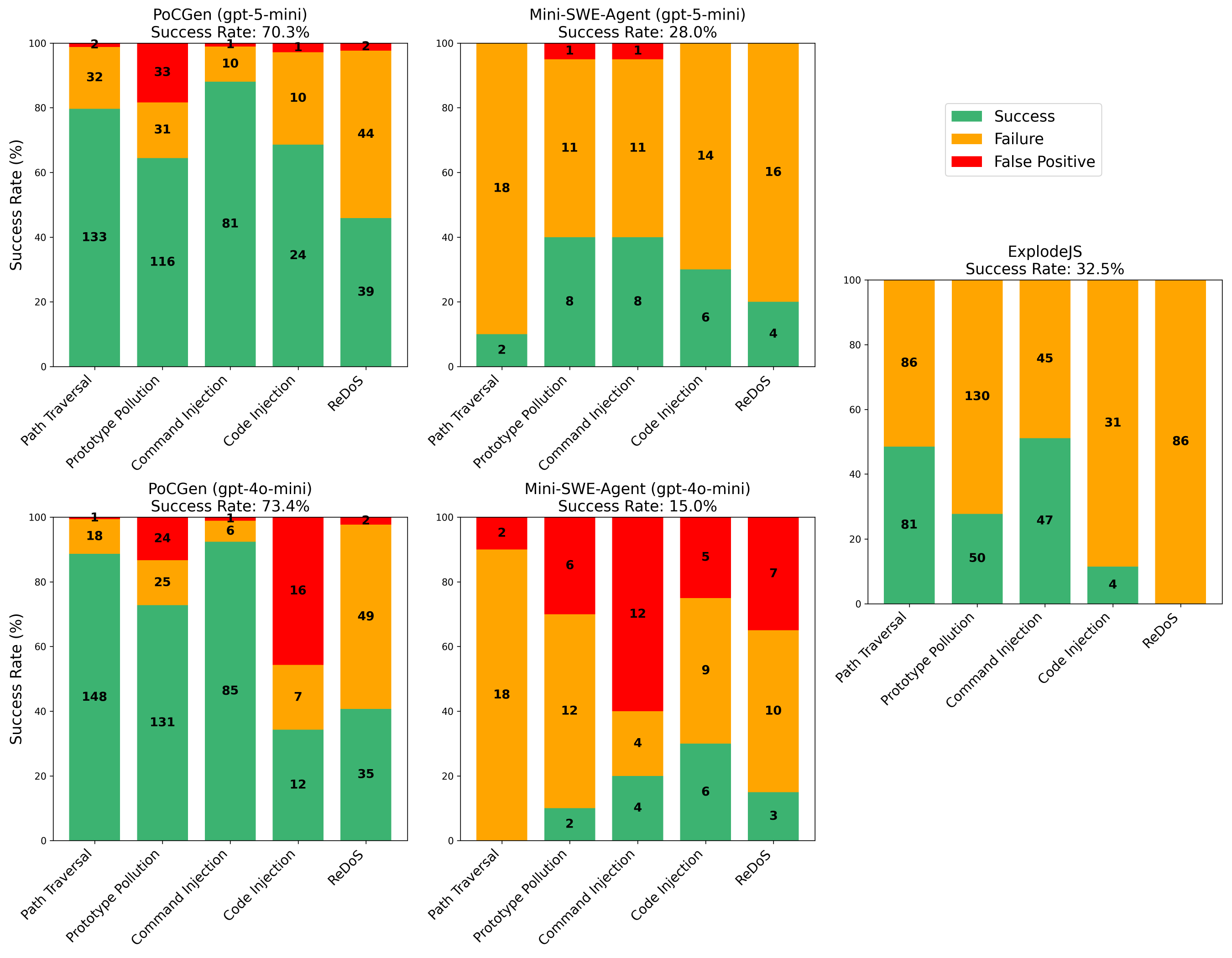}
  \caption{Effectiveness of \name{}, Mini-SWE-Agent, and Explode.js on SecBench.js.}
  \label{fig:eval_all}
\end{figure*}

\Cref{fig:eval_all} shows the effectiveness of \name{}, Mini-SWE-Agent, and Explode.js on the SecBench.js dataset.
\name{} successfully generates PoC exploits for 395 out of \numSecBenchTotal{} vulnerabilities, on average, which is 71\% of the vulnerabilities.
The standard deviation of the number of generated exploits across the three repeated runs is 4.5, amounting to 0.8\% of the dataset size.
Explode.js successfully generates PoC exploits for 182 out of \numSecBenchTotal{} vulnerabilities, which is 32\% of the vulnerabilities.
Mini-SWE-Agent generates successful PoC exploits for 28 out of the 100 vulnerabilities on average, with a standard deviation of 2.65.
The results show that \name{} outperforms Explode.js by 38 and Mini-SWE-Agent by 42 absolute percentage points.

When comparing the effectiveness of \name{} on different vulnerability types, we find that \name{} performs best on path traversal, and command injections, with success rates of above 79\%.
In case of Explode.js, it also performs better on path traversal and command injection compared to the other vulnerability types, with success rates between 50\% and 60\%.
Since Explode.js does not support ReDoS vulnerabilities, its success rate on ReDoS vulnerabilities is 0\%.
Mini-SWE-Agent succeeds in 40\% of cases on prototype pollution and command injection, while only succeeding in 10\% of cases on path traversal vulnerabilities.

We also evaluate \name{} and Mini-SWE-Agent using the \code{gpt-4o-mini} model, which is an older and less capable model than \code{gpt-5-mini}. 
The results show that \name{} maintains a high success rate, while Mini-SWE-Agent's effectiveness significantly drops by 13 percentage points, with an increase in false positives.

By investigating the reasons that \name{} fails to generate a successful PoC exploit, we observe that for 4\% of the vulnerabilities, the approach runs through 30 refinement iterations without generating a working exploit.
For 3.6\% of vulnerabilities, the token limits we set are reached, and the approach stops.
The packages in 1.2\% of the vulnerabilities do not export any functions.
Finally, for 1.2\% of vulnerabilities, \name{} itself evaluates its own generated exploits as false positives.

\subsection{RQ2: Ablation Study}

To evaluate the impact of each component on the effectiveness of \name{}, we perform an ablation study on the SecBench.js dataset.
We evaluate the following configurations of \name{} to measure the impact of each component on the success rate of generating PoC exploits.
\begin{itemize}
  \item \name{}: The complete \name{} approach with all components.
  \item noTaintPath: \name{} without the taint path analysis, described in~\cref{sec:taint_path}. In this configuration, no taint analysis is done, which means the approach does not utilize taint paths to find the vulnerable source, the LLM prompt does not contain the taint path, and also both of the context refiners, i.e., the body refiner and the missing declaration refiner, that depend on the taint path are not used.
  \item noFewShot: \name{} without the few-shot examples of exploits of similar vulnerabilities (\cref{sec:exploit_generation}).
  \item noUsageSnippet: \name{} without the usage snippet examples (\cref{sec:usage_snippets}).
  \item noDebuggerRefiner: \name{} without the debugger refiner (\cref{sec:runtime_refiners}).
  \item noErrorRefiner: \name{} without the error refiner (\cref{sec:runtime_refiners}).
\end{itemize}

\begin{table}
    \centering
    \caption{The effect of each component of \name{} on successfully generating PoC exploits.}
    \label{tab:ablation}
    \begin{tabular}{lrr}
        \toprule
        Configuration     & Valid Exploits & Success Rate \\
        \midrule
        \textbf{\name{}}  & \textbf{395}   & \textbf{71\%} \\
        noTaintPath       & 115            & 21\%         \\
        noUsageSnippet    & 196            & 35\%         \\
        noFewShot         & 375            & 67\%         \\
        noDebugger        & 377            & 67\%         \\
        noErrorRefiner    & 376            & 67\%         \\

        \bottomrule
    \end{tabular}
\end{table}

The success rates of these configurations on SecBench.js are shown in \cref{tab:ablation}.
These results show that all components contribute to \name{}'s overall effectiveness, with the taint analysis having the highest impact, followed by the usage snippet examples.

To better understand the contribution of iteratively refining the prompts, we also measure the number of refinement attempts required to generate a valid PoC exploit.
Iterative attempts for generating PoC exploits have a significant impact on the effectiveness of \name{}.
Only 27\% of the successful PoC exploits are generated in the first attempt, and 47\% are generated within two to ten refinements.
On average, successful PoC exploits are generated after 4 rounds of refinement.

We also analyze the component responsible for taint analysis, which has the highest impact on the effectiveness of \name{}.
The default CodeQL taint analysis is invoked for 72\% of the vulnerabilities, running 1.3 times per vulnerability on average.
In 39\% of these cases, the default taint analysis fails, and the approach falls back to our extended taint analysis.
Out of the extended taint analysis runs, 60\% still fail, which leads the approach to fallback to using the LLM to guess the vulnerable sinks, which still fails in 3\% of those cases.
The CodeQL taint analysis for remote sources, such as http request handlers, runs for 34\% of the vulnerabilities.

\subsection{RQ3: Costs}

To evaluate the costs of \name{}, we measure the time and token usage of the approach.
On average, each PoC exploit generation attempt takes 11 minutes.
Since successful attempts stop earlier than failing attempts, which require running the refiners, the successful runs complete on average in 7 minutes.
A significant portion of this time is spent on the CodeQL analyses, which take 51\% of the execution time.
The second-largest portion of the overall time is spent on querying the LLM, which takes 15\% of the overall time, on average.

We measure the number of tokens sent to and received from the LLM.
On average, for each attempt to generate an exploit, \name{} sends 32,059 tokens to the LLM and receives 8,644 tokens in response.
With the current pricing of the OpenAI API (as of February 2026), this costs \$0.025 per exploit generation attempt.
For the successful attempts, this cost is even lower, with only \$0.009 per vulnerability, on average.

\subsection{RQ4: Recent Real-World Vulnerability Reports}

To further evaluate the usefulness and generalizability of \name{}, we apply it to recent vulnerabilities reported on the GitHub Security Advisories database.
We select vulnerabilities that match one of our five vulnerability types and are reported between January 1, 2025, and January 8, 2026.
This results in a total of 183 vulnerability reports.
We then manually inspect the vulnerability reports and filter out vulnerabilities that are withdrawn, where the vulnerable package is no longer available on npm, where the vulnerable code is in a language other than JavaScript, or where exploiting the vulnerability requires a different operating system, setting up a specific server, or some other complex environment.
The final set contains 126 vulnerabilities, to which we apply \name{} to generate PoC exploits.
Since the knowledge cutoff date of \code{gpt-5-mini} is May 31, 2024, these vulnerabilities are not expected to be in the training data of the LLM, which allows us to evaluate the generalizability of \name{} to new vulnerabilities.

\name{} generates valid PoC exploits for 76 (60.3\%) of these vulnerability reports.
For six of the generated PoC exploits, the original vulnerability report did not have a PoC exploit.
For each of them, we create a pull request to add the generated PoC exploit to the vulnerability reports.
Five of the six pull requests have been accepted, and the generated exploits have been added to the corresponding vulnerability reports.
The remaining pull request is still under review.

For example, the popular package \code{tar-fs}, with over 25 million weekly downloads, had a path traversal vulnerability (CVE-2024-12905) reported on March 27, 2025.
This vulnerability affected three version ranges of this package, and before running \name{}, there were no public PoC exploits for this vulnerability.
\name{} generates a valid PoC exploit for this vulnerability, which is shown in \cref{fig:tarfs_exploit}.
To generate this exploit, \name{} uses eight refinement attempts, which include the error refiner and the debugger refiner.

\begin{figure}
  \begin{minted}{javascript}
const tarfs = require("tar-fs");
const fs = require("fs");
const path = require("path");

async function exploit() {
    // Create a writable stream to extract the tar content
    const extractStream = tarfs.extract('/', {
        // We can ignore the file type checks to allow the extraction of the malicious file
        ignore: (name) => false,
    });

    // Create a tar stream
    const tarStream = tarfs.pack().on('error', (err) => {
        throw err;
    });

    // Append the malicious entry to the tar stream
    tarStream.entry({ name: '/flag.txt', mode: 0o644 }, Buffer.from('This is a flag!'));

    // Finalize the tar stream
    tarStream.finalize();

    // Pipe the tar stream into the extract stream
    tarStream.pipe(extractStream);
}

// Call the exploit function
exploit();
  \end{minted}
  \caption{PoC exploit generated by \name{} for CVE-2024-12905 in the \code{tar-fs} package.}
  \label{fig:tarfs_exploit}
\end{figure}

Another example is our motivating example shown in \cref{fig:cve_no_poc_example}.
This vulnerability, reported on January 16, 2025 as \code{CVE-2024-36751}, is a regular expression denial of service (ReDoS) vulnerability in the \code{parse-uri} package.
As shown in \cref{fig:poc_exploit_ex1}, \name{} successfully generates a PoC exploit, which is now part of the vulnerability report on the GitHub Security Advisories database.

\subsection{RQ5: Qualitative Analysis}

\begin{figure}
  \centering
  \begin{minted}{javascript}
module.exports = function (seedPath, targetPath) {
  fsExtra.copySync(seedPath, targetPath);
  shell.exec('sh -c \'cd '+targetPath+' && npm install\'');
  shell.exec('sh -c \'cd '+targetPath+' && npm start\'');
}
  \end{minted}
  \caption{Example of a command injection vulnerability in the \code{n158} package.}
  \label{fig:n158}
\end{figure}

To better understand the strengths and limitations of \name{}, we present examples of successful and unsuccessful attempts by \name{} to generate PoC exploits.

\subsubsection{Successful Example: CVE-2023-26127}

The vulnerability report \code{CVE-2023-26127} describes a command injection vulnerability in the \code{n158} package.
This vulnerability is due to not sanitizing the input, which then flows to the \code{exec} function of \code{shelljs}.
However, the input is first passed to a copy function, as shown in \cref{fig:n158}.
This means that if the \code{seedPath} is set to a value where the corresponding path does not exist, the exploit fails before reaching the vulnerable sink.
This constraint is not easy to capture with symbolic execution, as attempted by Explode.js~\cite{Marques2025}, but using execution feedback and an LLM, \name{} is able to generate a working exploit.

In the first round of exploit generation, \name{} generates an exploit that calls \mintinline{javascript}{initProject( "someSeedPath", "targetPath; /usr/bin/genpoc")}.
After the execution of this exploit, the runtime feedback shows that there was a runtime error happening at the \code{fsExtra.copySync} function, which is due to the fact that no directory called \code{"someSeedPath"} exists.
The LLM then naively tries to fix this by changing \code{"someSeedPath"} to \code{"/path/to/valid/seed"}, which results in the same error.
It takes multiple refinements until the LLM requests the definition of \code{shelljs}, after which the LLM reasons that it needs to create the directory for a successful exploit.
This results in the final exploit shown in \cref{fig:example_success}.
\begin{figure}
  \centering
  \begin{minted}{javascript}
async function exploit() {
  const initProject = require("n158/src/cli/initProject");
  const seedPath = "/path/to/valid/seed";
  const targetPath = "targetPath; /usr/bin/genpoc";
  const fs = require('fs');
  if (!fs.existsSync(seedPath)) {
      fs.mkdirSync(seedPath, { recursive: true });
  }
  const result = await initProject(seedPath, targetPath);
}
await exploit();
  \end{minted}
  \caption{Example of a successful PoC exploit generation by \name{}.}
  \label{fig:example_success}
\end{figure}

\subsubsection{Unsuccessful Example: GHSA-3486-rvxc-hrrj}

The vulnerability report \code{GHSA-3486-rvxc-hrrj} describes a command injection vulnerability in the \code{gitblame} package.
The package exports one function, which takes a file path as an argument and after some processing passes it to \code{exec}.
The relevant parts of the source code are shown in \cref{fig:gitblame}.

\begin{figure}
  \centering
  \begin{minted}{javascript}
module.exports = function (file, cb) {
  var dirname = path.dirname(file);
  var filename = path.basename(file);
  exec('git blame ' + filename, {cwd: dirname}, ... );
}   
  \end{minted}
  \caption{Example of a command injection vulnerability in the \code{gitblame} package.}
  \label{fig:gitblame}
\end{figure}

\name{} is not able to generate a working exploit for this vulnerability.
The validator requires the command \code{/usr/bin/genpoc} to be executed, meaning the injected payload must contain a \code{/} character.
However, because the input is split by \code{path.basename} and \code{path.dirname} before reaching \code{exec}, any \code{/} in the payload is stripped, so no payload can simultaneously satisfy both the vulnerability trigger and the validator.
As a result, even with multiple refinements, \name{} is unable to generate a working exploit, although the exploit for this vulnerability in SecBench.js uses the simple payload "\code{\& touch gitblame}", which avoids this constraint entirely.

\section{Threats to Validity}
One threat to internal validity is the LLM's potential to recall exploits from its training data.
To mitigate this, we use the same LLM for the Mini-SWE-Agent baseline, which we would expect to perform similarly well as \name{} if the results were simply due to memorization.
Another threat is that the labeling process of false positive PoC exploits is done manually, which can introduce human bias.
To reduce the effects of this bias, we devise a set of objective criteria that precisely define what constitutes a false positive, reducing reliance on subjective judgment. For example, prototype pollution cases where the \code{Object.prototype} is part of an assignment or a function call are considered false positives.
A threat to external validity is that our evaluation is limited to five vulnerability types.
However, these vulnerability types correspond to the five most common threat classes in server-side JavaScript~\cite{Bhuiyan2023}.
Adapting the approach to other vulnerability types would require adapting our prompts, the sinks used by the taint analysis, and the validators.
Another threat to external validity is that we evaluate our approach primarily on the SecBench.js dataset, which may not be representative of all vulnerabilities in the npm ecosystem.
To mitigate this threat and the internal validity threat of model memorization, we also evaluate \name{} on 126 vulnerabilities in the npm ecosystem that were reported since January 2025, i.e., after the knowledge cutoff of the LLMs we use.
Our approach is implemented for JavaScript and the npm ecosystem, which may limit its applicability to other programming languages and ecosystems.
We believe that the core ideas of \name{}, i.e., how to combine LLMs with program analysis techniques, could be adapted to other contexts, but leave this for future work.

\section{Related Work}

\paragraph{Vulnerability Detection}
Greybox fuzzing~\cite{afl2013,Boehme2019}, applied to source code~\cite{Sherman2025} or binaries~\cite{Dinesh2020}, is a common approach for vulnerability detection.
To evaluate fuzzing, techniques for reverting fixes~\cite{Zhang2022} and benchmarking methodologies~\cite{Klees2018,Li2021} have been proposed.
Learning-based vulnerability detection includes neural classification~\cite{Li2018a,Fu2022,Chakraborty2022}, graph neural networks~\cite{Li2021c,Steenhoek2024}, and combinations of LLMs with static analysis.
Similar to our work, the latter leverages CodeQL to identify taint flows~\cite{Li2024}.
To support learning-based detection, datasets from commit histories~\cite{Zheng2021} and large-scale vulnerability generation approaches~\cite{fse2022-vuln,icse2023-VulGen} have been introduced.
Vulnerability detection is orthogonal to our work, as we assume vulnerabilities are already described in a report but lack an exploit.

\paragraph{Detecting and Exploiting Node.js Vulnerabilities}
The closest work to \name{} is Explode.js~\cite{Marques2025}, which finds vulnerabilities and generates PoC exploits for npm packages.
Explode.js uses static dataflow analysis to extract the sequence of function calls required to propagate attacker input to a vulnerable sink.
It then applies symbolic execution and constraint solving to generate a PoC exploit.
However, Explode.js does not model external functions and libraries during symbolic execution, which limits its effectiveness, as the npm ecosystem heavily relies on small, reusable packages.
\name{} addresses this limitation by leveraging LLMs to generate exploits and by incorporating runtime feedback.
LLMs, trained on large code corpora, can better predict the behavior of external functions and reason about inputs that exercise specific program paths.

Other approaches have used symbolic execution to generate exploits for vulnerabilities.
FAST~\cite{Kang2023JSAI} applies bi-directional dataflow analysis to detect taint paths efficiently, enabling scalable vulnerability detection.
It generates exploits by concretizing symbolic path constraints once a vulnerability is found.
Node-Medic~\cite{Cassel2023} and Node-Medic-FINE~\cite{Cassel2025} combine dynamic taint analysis with symbolic execution to detect and generate exploits for Node.js packages.
Node-Medic-FINE further incorporates fuzzing to generate inputs and explore additional execution paths.
All three approaches are outperformed by Explode.js~\cite{Marques2025}, which we therefore use as a baseline in our evaluation.

\paragraph{Test Generation for Security Vulnerabilities}
Zhang et al.~\cite{Zhang2023LLMSecTests} and Gao et al.~\cite{Gao2025vuln} use LLMs to generate unit tests for Java vulnerabilities given a PoC exploit.
Their goal is to encourage developers to update vulnerable dependencies and prevent supply chain attacks.
In contrast, our work generates code that directly exploits a vulnerable package, rather than exploiting it through a third-party dependency.

\paragraph{LLM-Assisted Attacks}
Recent work has explored the potential of LLMs for attacking vulnerable software.
PwnGPT~\cite{Peng2025} uses LLMs with reasoning models to generate exploits for capture-the-flag (CTF) challenges.
PentestGPT~\cite{Deng2024} uses LLMs for penetration testing.
Both of these approaches focus on system-level penetration tasks, which differs from the more isolated task of providing a PoC exploit for an individual package.
Moreover, these approaches do not enable the LLM to directly benefit from sophisticated tools, e.g., for statically analyzing code or interactively debugging a candidate exploit.
Xu et al.~\cite{Xu2024auto} developed an LLM agent with command-line access to exploit vulnerabilities in Linux and Windows applications.
Charan et al.~\cite{Charan2023} investigated using LLMs to generate payloads for exploiting vulnerabilities.

\paragraph{Vulnerability Mitigation and Repair}
Mitigating vulnerabilities can involve removing unused dependencies~\cite{Koishybayev2020} or reducing the privileges of vulnerable code~\cite{ccs2021mir}.
Repairing vulnerabilities can be achieved by fine-tuning LLMs to find fixes~\cite{Berabi2024}, using LLM agents~\cite{Zhang2024c}, or applying generative adversarial networks (GANs)~\cite{Harer2018}.

\paragraph{JavaScript and Npm Ecosystem Security}
The npm ecosystem faces various security issues, such as injection attacks~\cite{ndss2018}, ReDoS~\cite{usenixSec2018}, and malicious packages~\cite{Sejfia2022}.
Several empirical studies have analyzed npm from a security perspective, including vulnerability propagation~\cite{Zimmermann2019,Liu2022} and how developers address vulnerabilities~\cite{Zahan2022weak}.
Householder et al.~\cite{Householder2020} found that most vulnerability reports lack a public PoC exploit for at least one year.
Yadmani et al.~\cite{Yadmani2023} further showed that many PoC exploits on GitHub are themselves malicious.
These findings motivate our work on automated PoC exploit generation.
To support further research, Bhuiyan et al.~\cite{Bhuiyan2023} introduced the SecBench.js dataset, and Brito et al.~\cite{Brito2023} created the VulcaN dataset.

\section{Ethical Considerations}
Automating the generation of exploits raises ethical concerns, as it could potentially be misused by malicious actors to create exploits for vulnerabilities that have not yet been patched.
Since our approach targets "proof-of-concept" exploits, which are typically used for testing and demonstration purposes, the generated exploits may not be directly usable for malicious purposes.
Moreover, we believe that our approach can benefit the developers of vulnerable packages and the security community in general by automating part of the vulnerability management process, which can lead to faster patching and improved security.

Our evaluations have all been on public vulnerability reports, which have already been responsibly disclosed, and for which exploits may already exist.
Furthermore, as our datasets consist of vulnerable packages, we were able to run all the exploits on isolated environments without risking harm to real users.
Finally, we have made the source code of \name{} and the datasets used in our evaluation publicly available to encourage further research in this area.

\section{Conclusion}
This paper presents \name{}, an LLM-based approach to automatically generate proof-of-concept exploits for vulnerabilities in npm packages.
\name{} extracts information from the vulnerability report and the codebase to generate a PoC exploit using an LLM.
The generated PoC exploits are validated using a set of runtime checkers, and the prompt is refined using static and dynamic information to generate a valid exploit.
We evaluated \name{} on a dataset of \numSecBenchTotal{} vulnerabilities in npm packages.
\name{} generates PoC exploits for 71\% of these vulnerabilities, outperforming the state-of-the-art approaches Explode.js and Mini-SWE-Agent by 38 and 42 absolute percentage points, respectively.
\name{} also generates PoC exploits for 60\% of recently reported vulnerabilities from the GitHub Security Advisories database.

By automating the generation of PoC exploits, \name{} enables developers and security teams to more rapidly understand and address vulnerabilities, reducing the time between vulnerability disclosure and patch deployment.
This automation also improves regression testing, as the generated PoC exploits can be directly used to verify the effectiveness of fixes and to prevent vulnerabilities from reappearing in the future.
For security researchers, \name{} provides an automated way to evaluate the effectiveness of existing mitigation strategies across large sets of vulnerabilities.
Furthermore, \name{} can improve the quality of existing vulnerability reports, including those that are poorly documented or lack existing exploits.
Finally, automated PoC generation can facilitate responsible vulnerability disclosure by providing clear, actionable evidence to affected parties, encouraging timely remediation.

\section*{Data Availability}
\label{sec:data}
The source code of \name{}, the new dataset, and all experiment scripts are available at \url{https://github.com/sola-st/PoCGen}, and also as a permanent snapshot at \url{https://doi.org/10.5281/zenodo.19550952}.

\begin{acks}
This work was supported by the German Research Foundation (DFG; projects 492507603, 516334526, and 526259073). 
\end{acks}

\bibliographystyle{ACM-Reference-Format}
\bibliography{referencesMichael,references}

\end{document}